\pdfoutput=1

\documentclass[11pt]{article}

\usepackage[preprint]{acl}

\usepackage{times}
\usepackage{latexsym}
\usepackage{subcaption} 
\usepackage{booktabs}
\usepackage{framed}
\usepackage{multirow}
\usepackage{amsmath, amssymb}
\usepackage{caption}
\usepackage[T1]{fontenc}

\usepackage[utf8]{inputenc}

\usepackage{microtype}

\usepackage{inconsolata}

\usepackage{graphicx}
\usepackage{float}
\usepackage{stfloats}

\title{LLM as Explainable Re-Ranker for Recommendation System}

\author{
  *Yaqi Wang \\
  Carnegie Mellon University \\
  \texttt{yaqiwang@andrew.cmu.edu} \\\And
  *Haojia Sun \\
  Carnegie Mellon University \\
  \texttt{haojias@andrew.cmu.edu} \\\And
  *Shuting Zhang \\
  Carnegie Mellon University \\
  \texttt{shuting2@andrew.cmu.edu} \\}

\begin{document}
\maketitle
\begin{abstract}
The application of large language models (LLMs) in recommendation systems has recently gained traction. Traditional recommendation systems often lack explainability and suffer from issues such as popularity bias. Previous research has also indicated that LLMs, when used as standalone predictors, fail to achieve accuracy comparable to traditional models. To address these challenges, we propose to use LLM as an explainable re-ranker, a hybrid approach that combines traditional recommendation models with LLMs to enhance both accuracy and interpretability. We constructed a dataset to train the re-ranker LLM and evaluated the alignment between the generated dataset and human expectations. Leveraging a two-stage training process, our model significantly improved NDCG, a key ranking metric. Moreover, the re-ranker outperformed a zero-shot baseline in ranking accuracy and interpretability. These results highlight the potential of integrating traditional recommendation models with LLMs to address limitations in existing systems and pave the way for more explainable and fair recommendation frameworks. Code is available at \url{https://github.com/Arlene036/LLM4FairRec}.

\end{abstract}

\section{Introduction}
Recommendation systems have become indispensable in internet applications, delivering personalized suggestions to users and significantly enhancing their experience. These systems are widely employed across various domains, including e-commerce~\cite{rec_e-commerce}, streaming services~\cite{rec_stream}, and social media platforms~\cite{rec_social}. However, despite their growing adoption, recommendation systems face critical limitations in terms of explainability~\cite{rec_explain} and fairness~\cite{rec_fairness}. Traditional systems often prioritize popular items over personalized needs~\cite{popularity_bias}, resulting in biased recommendations that fail to cater to individual user preferences. Furthermore, these systems typically lack transparent and trustworthy explanations for their recommendations, making it difficult for users to understand or trust the results.

One of the primary challenges in recommendation systems is the lack of explainability~\cite{rec_explain}. Existing methods tend to focus predominantly on improving recommendation accuracy, while the generation of user-friendly explanations remains an underexplored area. This lack of transparency undermines user trust and limits the effectiveness of recommendations. Another significant issue is position bias~\cite{position_bias}, particularly in systems employing large language models (LLMs). LLMs tend to allocate varying attention weights to tokens based on their positions, leading to unfair rankings. Additionally, traditional systems are often plagued by popularity bias~\cite{popularity_bias}, where popular items are overly favored, neglecting niche or personalized preferences, which ultimately fails to align with individual user interests.

To address these challenges, we propose a novel framework that incorporates LLMs as re-rankers to enhance the explainability of recommendation results and mitigate biases. By integrating LLMs into the ranking phase, our approach generates user-readable explanations that improve transparency and user trust. To reduce position bias, we introduce a bootstrapping mechanism that shuffles candidate lists, ensuring that the LLM's attention is distributed more fairly. Moreover, we employ Direct Preference Optimization (DPO) and Reward-Preference Optimization (RPO) to refine ranking quality and better capture user preferences, thereby addressing both position and popularity biases.

Our contributions are as follows:
\begin{itemize}
    \item \textbf{Integration of LLMs and Recommendation Models:} We incorporate LLMs into the ranking phase of recommendation systems, enabling the generation of user-readable explanations that enhance transparency and trust.
    \item \textbf{Two-Stage Training Framework:} We propose a two-stage training approach that combines Supervised Fine-Tuning (SFT) and Direct Preference Optimization (DPO). This framework mimics the optimization logic used in traditional systems for handling positive and negative samples, improving ranking accuracy and explanation quality.
    \item \textbf{Bootstrapping Mechanism:} To mitigate position bias inherent in LLMs, we introduce a bootstrapping method that randomizes candidate list orders, ensuring fairer attention distribution and unbiased rankings.
    \item \textbf{Diversified Candidate Generation:} Our framework enhances candidate list diversity by integrating collaborative filtering, knowledge graph-based methods, and random retrieval, ensuring a rich and varied pool of recommendation candidates.
\end{itemize}

By addressing the limitations of existing recommendation systems and introducing these innovations, our framework not only improves recommendation accuracy and fairness but also enhances user trust and satisfaction through transparent and interpretable results. This paper details our proposed framework, its methodology, and experimental validation to demonstrate its efficacy in real-world scenarios.

\section{Related Work}
The application of Large Language Models (LLMs) in recommendation systems has rapidly advanced due to their semantic understanding and reasoning capabilities \cite{wu2024surveylargelanguagemodels}. Current research primarily explores two areas: LLMs with textual data and those integrated with structured graph information. In text-based approaches, LLMs demonstrate potential in generating flexible and explainable recommendations but face challenges with order-sensitive tasks and domain-specific nuances. For instance, while ChatGPT shows coherence in zero-shot ranking and explanation generation, it struggles with sequential recommendation tasks due to positional biases \cite{liu2023chatgptgoodrecommenderpreliminary}. Frameworks like TALLRec employ instruction tuning and LoRA to align LLMs with recommendation prompts, improving efficiency but limiting flexibility in complex scenarios \cite{10.1145/3604915.3608857}. LLM-based Zero-Shot Rankers frame ranking as a language task but encounter difficulties in capturing nuanced user-item interactions \cite{hou2024largelanguagemodelszeroshot}.

Graph-based approaches leverage LLMs for enhanced context, reasoning, and explanation capabilities. As enhancers, models like GaCLLM integrate multi-hop neighbor information to enrich user-item profiles, though scalability remains an issue \cite{du2024largelanguagemodelgraph}. When used for reasoning, LLMs, such as in LLMRG and GLRec, construct causal reasoning graphs or encode high-order interactions, improving interpretability but increasing computational costs \cite{LLM_graph_reasoning, wu2023exploringlargelanguagemodel}. For explanations, models like PEARLM and subgraph-based methods enhance transparency using knowledge graphs, yet they face adaptability challenges in dynamic or evolving domains \cite{balloccu2024faithfulpathlanguagemodeling, shi2024llmpoweredexplanationsunravelingrecommendations}.

\section{Methodology}


\subsection{Dataset Construction}

\begin{figure*}[t]
  \centering
\includegraphics[width=0.9\textwidth]{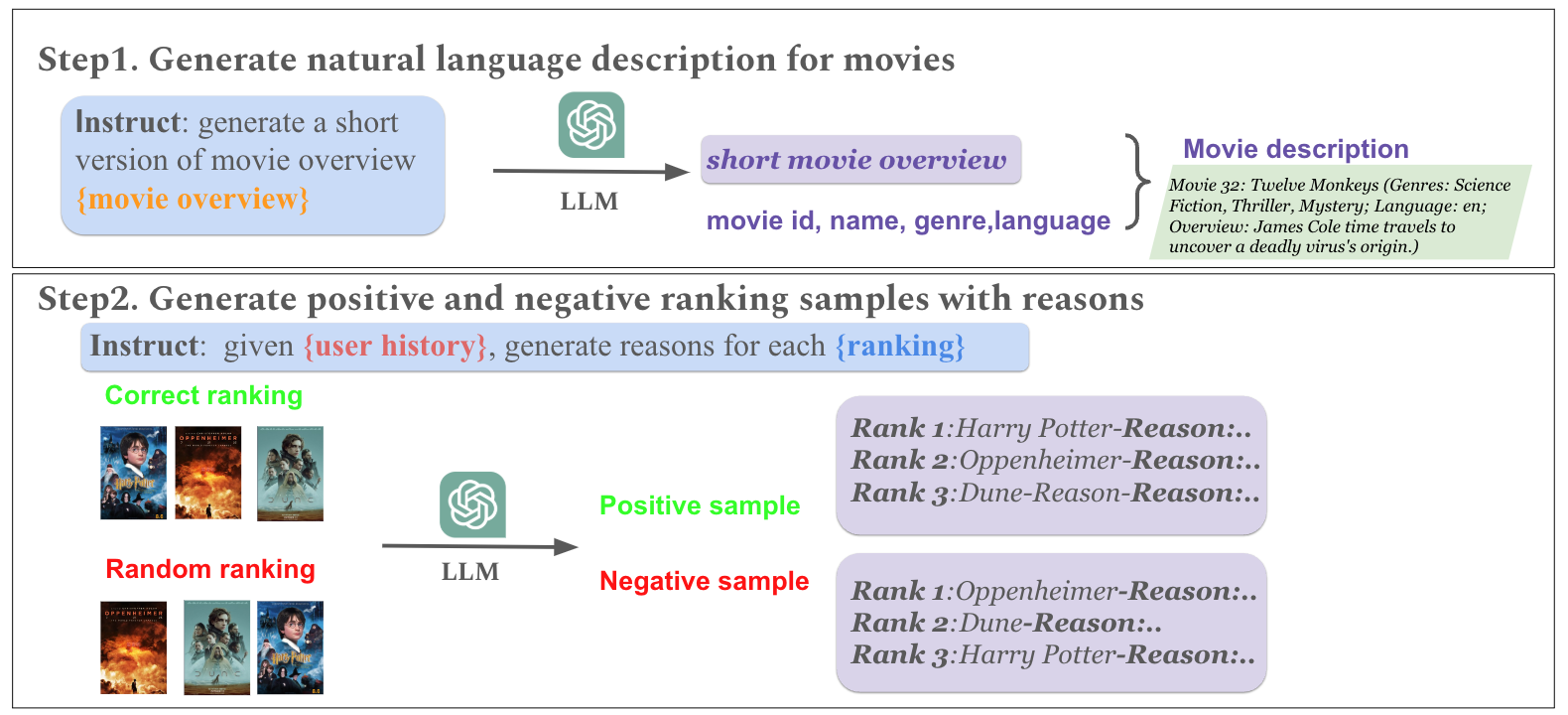}
  \caption{Design of Data Extraction.}
  \label{fig:data_creation}
\end{figure*}

To enable the re-ranker to generate accurate rankings and compelling explanations, we constructed a dataset in two stages. This process combines movie metadata, user histories, and ranking explanations to provide both positive and negative training samples for the model, as shown in Figure~\ref{fig:data_creation}.

First, we used GPT-4-o-mini to summarize the overviews to less than 15 words for each movie. Combining these short overviews with \textit{movie id}, \textit{name}, \textit{genre}, and \textit{language}, we created natural language descriptions for all movies, ensuring compatibility with LLM input constraints.

Next, we generated ranking samples to train the reranker. Positive samples were constructed by providing GPT-4-o-mini with user histories and the correct ranking of movies, prompting it to generate reasons for the ranking order. Negative samples were created by shuffling the rankings randomly and generating reasons for the incorrect order. These samples were used to fine-tune the re-ranker through Supervised Fine-Tuning (SFT) and Direct Preference Optimization (DPO), allowing the model to learn to differentiate between correct and incorrect rankings while producing user-aligned explanations.

\subsection{Training}
Our training phase consists of two distinct stages: Supervised Fine-Tuning (SFT) and Direct Preference Optimization (DPO). Each stage contributes to optimizing the LLM for ranking and explanation generation, ensuring both accuracy and interpretability in recommendation tasks.

\begin{figure}[t]
  \includegraphics[width=0.5\textwidth]{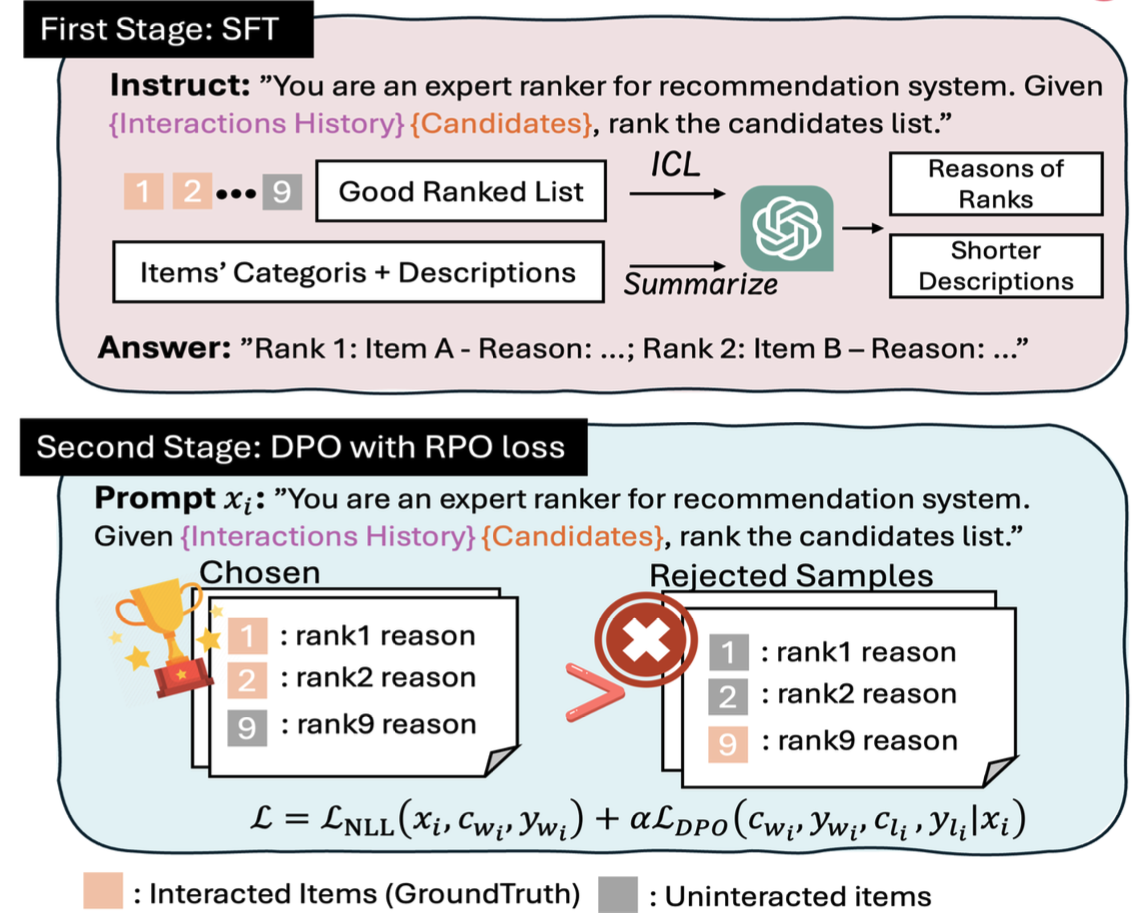}
  \caption{Training Pipeline}
  \label{fig:train}
\end{figure}

The primary objective of the training phase is to equip the LLM with robust ranking and explanation generation capabilities. The first stage, Supervised Fine-Tuning (SFT), focuses on introducing the foundational skills required for these tasks by using structured prompts and labeled data to align the model’s outputs with human-annotated Ground Truth rankings. The second stage, Direct Preference Optimization (DPO)~\cite{DPO}, builds on this foundation by refining the model’s ability to distinguish between positive (chosen) and negative (rejected) samples, thereby enhancing its ranking accuracy and user preference alignment. Together, these stages ensure the LLM can generate both accurate rankings and clear, user-friendly explanations.

\subsubsection{Stage 1: Supervised Fine-Tuning (SFT)}
The first stage, Supervised Fine-Tuning (SFT), aims to train the LLM to learn fundamental ranking and explanation generation capabilities through supervised learning. Using structured prompts and labeled data, the model begins to understand user preferences and produce interpretable ranking results. 

To further enhance the model’s reasoning capabilities, we employ Noisy Layer Fine-Tuning (NEF)~\cite{NEF_tune}, a method that introduces controlled noise into each model layer during training. NEF  ensures that the model remains robust across different user profiles and item categories, preventing overfitting to specific patterns in training data. This technique improves the model’s robustness and generalization, ensuring better performance in both ranking and explanation tasks.

By the end of SFT, the model generates ranked lists accompanied by detailed explanations, such as: 

\begin{framed}
    \noindent User 5: \\
    Rank 1: Movie A - Highly relevant due to user preference for Romance and strong critical reviews. \\
    Rank 9: Movie B - Less relevant as it belongs to genres not preferred by the user.
\end{framed}
\noindent These outputs represent the model’s foundational understanding of user preferences and ranking logic.

\subsubsection{Stage 2: Direct Preference Optimization}
Building on the foundation laid by SFT, the second stage, Direct Preference Optimization (DPO)~\cite{DPO}, refines the LLM’s ability to prioritize correct rankings. This stage focuses on optimizing the model’s preference for positive (chosen) samples over negative (rejected) samples. Positive samples, or chosen samples, are derived from the ranking results generated during the SFT phase. These represent desirable ranking orders paired with explanations. Negative samples, or rejected samples, are created by shuffling the ranking order of the chosen samples, introducing incorrect rankings that the model must learn to reject.

The training process involves presenting the model with input prompts alongside paired ranking outputs—one chosen and one rejected. The LLM is optimized to score the chosen sample higher than the rejected one, thereby maximizing the score margin between positive and negative samples. This optimization process is guided by a combination of loss functions. First, the DPO Loss maximizes the margin between chosen and rejected scores, ensuring the model consistently favors correct rankings.

Additionally, we incorporate RPO Loss~\cite{RPO}, which refines the model’s preference modeling by applying weighted adjustments to the SFT and DPO losses. The RPO Loss builds on the principles of iterative preference modeling. Unlike DPO, which focuses solely on maximizing score margins, RPO introduces a dynamic weighting mechanism that integrates the strengths of both supervised fine-tuning and margin-based optimization. Specifically, RPO assigns higher weights to chosen samples that align strongly with SFT objectives while penalizing rejected samples more effectively. This ensures that the model does not only prioritize correct rankings but also retains the ability to generate high-quality explanations consistent with supervised learning goals. By iteratively refining preferences, RPO mitigates overfitting to either component and promotes balanced optimization across ranking accuracy and explanation clarity. This approach has been shown to enhance the robustness of LLMs in recommendation tasks involving noisy or complex inputs~\cite{RPO}.

\subsection{Inference}

\begin{figure*}[t]
\centering
  \includegraphics[width=0.8\textwidth]{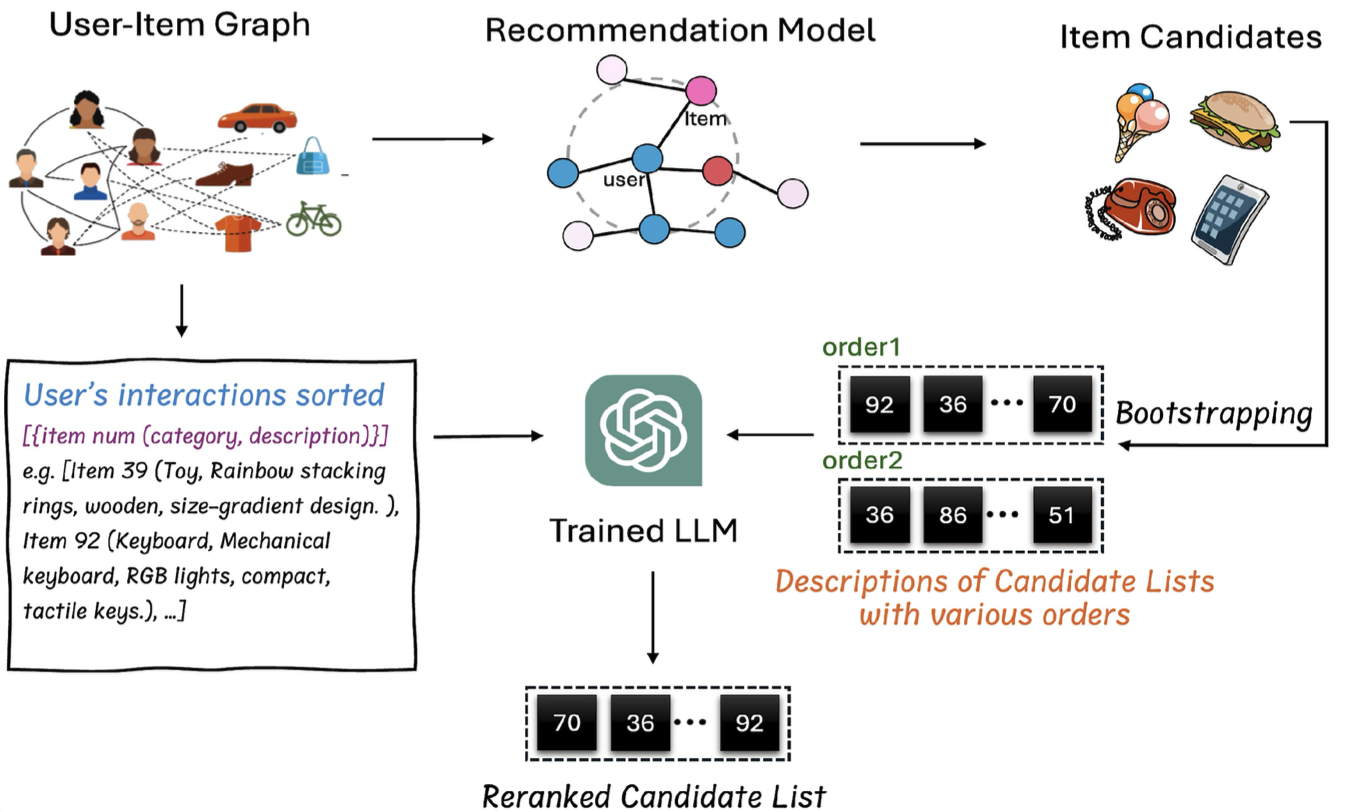}
  \caption{LLM as Re-Ranker}
  \label{fig:inference}
\end{figure*}

After training, the model is utilized for inference, with the goal of recommending $N$ items to a specific user. As shown in Figure~\ref{fig:inference}, the input to the LLM consists of two components: the user's historical interactions (sampled to 10 entries) and a list of 15 candidate items. The LLM is tasked with re-ranking this candidate list.

To address potential biases, we implement several strategies. First, to mitigate popularity bias—where popular items tend to dominate the top positions—we shuffle the candidate list before feeding it into the LLM. Second, to counteract the positional bias introduced by the LLM itself (as it may assign varying importance to tokens depending on their positions), we employ bootstrapping to generate multiple candidate lists with different orders. 

The shuffled candidate lists, along with the user's interaction history, are fed into the LLM to produce multiple ranked outputs. A self-consistency module is then applied to determine the final ranking. Specifically, we calculate the position index for each item across all ranked outputs and sum these indices. Items are then ordered based on their cumulative position scores, with lower scores being ranked higher. This approach ensures a robust and unbiased final ranking.

\section{Experiments}
%
In this section, we will first prove that our LLM-generated data could align with human expectations, and then we will compare our trained model with other baseline models on three different datasets and analyze the results.

\subsection{Experiments Setup}
In our experiments, we evaluate the performance of the proposed framework using candidate lists generated by three methods: random sampling (Randomly Retrieved), collaborative filtering based on matrix factorization (SVD), and knowledge graph-based recommendations using RippleNet. The comparison includes two baseline models: no re-ranker (None Ranker), a zero-shot re-ranker, to compare with our trained re-ranker utilizing the SFT-DPO pipeline. To evaluate the performance of each model, for each recommendation method, we construct a dataset consisting of 1000 samples. Each sample includes the top-15 items predicted by the recommendation model, accompanied by their natural language descriptions. Both the baseline LLM and the trained LLM perform inference on these 1000 samples. Finally, a parser processes the LLM outputs to compute the relevant evaluation metrics.

For both the training and inference phases, a single NVIDIA A100 GPU was utilized. The unsloth \cite{unsloth} framework was employed to accelerate both training and inference processes, with Mistral-7b-Instruct serving as the base LLM for training. During inference, the number of bootstrapping iterations was fixed at three to maintain a controlled budget.

\subsection{Human Evaluation of LLM-Generated Dataset}
We manually curated 30 entries as a control dataset, with two team members each contributing 15 samples. To evaluate the quality of the LLM-generated dataset, we recruited five peers from the same field. Each participant was provided with 10 manually created entries and 10 LLM-generated entries, randomly sampled. They were asked to rate the quality of each entry on a scale from 1 to 5, based on the persuasiveness of the ranking.

To assess the alignment of the LLM-generated data with human preferences, we conducted a t-test with the null hypothesis being that the quality of the manually created data is equal to that of the LLM-generated data. The experiment yielded a t-value of 1.24 and a p-value of 0.22. As the p-value indicates no significant difference, we conclude that the LLM-generated dataset aligns well with human expectations. Thus, our constructed dataset meets the desired quality criteria.

\subsection{Metrics}
We aim to evaluate the model's ranking performance and explainability, i.e., the quality of generated ranking reasons. 
\subsubsection{Ranking Performance} To assess performance of ranking, we use standard ranking metrics: Hit Ratio, Recall, Precision, and NDCG (Normalized Discounted Cumulative Gain). 
\paragraph{Hit Ratio (HR)}  
Hit Ratio measures whether the ground truth item appears in the top-\(N\) recommendations for a given user. It is defined as:
\[
HR@N = \frac{1}{|U|} \sum_{u \in U} \mathbb{I} \big( \text{rank}(i_u) \leq N \big)
\]
where \(U\) is the set of users, \(i_u\) is the ground truth item for user \(u\), \(\text{rank}(i_u)\) is the rank position of \(i_u\) in the recommendation list, and \(\mathbb{I}(\cdot)\) is an indicator function that equals 1 if the condition is true and 0 otherwise.

\paragraph{Recall}  
Recall measures the proportion of relevant items retrieved in the top-\(N\) recommendations. 
\[
Recall@N = \frac{\text{Number of relevant items in top-}N}{\text{Total number of relevant items}}
\]

\paragraph{Precision}  
Precision indicates the fraction of recommended items in the top-\(N\) that are relevant.
\[
Precision@N = \frac{\text{Number of relevant items in top-}N}{N}
\]
This metric reflects the accuracy of the recommendations within the top-\(N\) ranked items.

\paragraph{NDCG}  
NDCG (Normalized Discounted Cumulative Gain) evaluates the ranking quality by considering both the relevance and the position of the recommended items. It is computed as the ratio of the Discounted Cumulative Gain (DCG) to the Ideal Discounted Cumulative Gain (IDCG), which normalizes the score to a range between 0 and 1:

\[
NDCG@N = \frac{DCG@N}{IDCG@N}
\]

The Discounted Cumulative Gain (DCG) is defined as:
\[
DCG@N = \sum_{i=1}^{N} \frac{\mathbb{I}(i \text{ is relevant})}{\log_2(i+1)}
\]

The Ideal Discounted Cumulative Gain (IDCG) represents the maximum possible DCG for a perfect ranking:
\[
IDCG@N = \sum_{i=1}^{|R_u|} \frac{1}{\log_2(i+1)}
\]

where \(R_u\) is the ordered list of relevant items for user \(u\). This formulation ensures that NDCG reflects both the relevance of recommended items and their positions within the ranked list.

To further validate the significance of the observed improvements, we conducted a paired t-test to compare the performance of the trained Re-Ranker models (SFT-DPO-Trained) against both the Zero-Shot Re-Ranker and the None Ranker baselines. The t-test was performed on key metrics such as NDCG, Hit Ratio, and Precision across all Top-N settings (Top 3, Top 5, and Top 10), indicating to test whether the trained models consistently achieve statistically significant improvements ($p \le 0.05 $) over the baselines

\subsubsection{Explainability}  
To evaluate the explainability of the Re-Ranker
system, we leverage human evaluation to evaluate
the quality of generated reasons for ranking. Participants were presented with ranked lists generated by the SFT-DPO model and the zero-shot ranking model. For each list, participants were provided with the corresponding natural language explanations for the ranking. They were asked to rate the persuasiveness and clarity of the explanations on a 5-point Likert scale, where 1 indicates "not persuasive at all" and 5 indicates "very persuasive."

\subsection{Results}
As shown in Table~\ref{tab:all_results}, the LLM Re-Ranker exhibits notable improvements across multiple metrics, particularly when applied to weaker base recommendation models. For randomly retrieved candidates (Table~\ref{tab:random_candidates}), the SFT-DPO model significantly outperforms both the non-ranker baseline and the zero-shot LLM. Notable gains are observed in metrics like NDCG@3, NDCG@5, and NDCG@10, with p-values consistently below the 0.05 threshold, demonstrating statistical significance. Similarly, for item-based collaborative filtering (Table~\ref{tab:item_based}), the LLM Re-Ranker enhances Recall@5, Precision@5, and NDCG@10, with improvements becoming significant in cases where the baseline model performance is relatively weak. However, the performance gains diminish for stronger base models, such as knowledge graph-based systems (Table~\ref{tab:knowledge_graph}). While SFT-DPO achieves marginal improvements in NDCG@10, the differences in most metrics are not statistically significant. This suggests a diminishing return when richer contextual information is already incorporated into the base recommendation model.

Overall, the results highlight the effectiveness of the LLM Re-Ranker in elevating weaker recommendation models but indicate challenges in leveraging its full potential when applied to already robust systems. This underscores the importance of selecting an appropriate base recommendation model to maximize the benefits of re-ranking with an LLM.

\begin{table*}[!ht]
\caption{Performance Comparison of Different Recommendation and Re-Ranker Models}
\label{tab:all_results}
\centering
\small 
\begin{subtable}{\textwidth}
\centering
\caption{Randomly Retrieved Candidates}
\label{tab:random_candidates}
\begin{tabular}{lccccc}
\toprule
\multirow{2}{*}{\textbf{Metrics}} & \multirow{2}{*}{\textbf{Non-Ranker}} & \multirow{2}{*}{\textbf{Zero-shot}} & \multirow{2}{*}{\textbf{SFT-DPO}} & \multicolumn{2}{c}{\textbf{p-value}} \\
\cmidrule[\lightrulewidth](lr){5-6}
& & & & \textbf{Zero vs Non} & \textbf{SFT vs Non} \\
\midrule
Hit Ratio@3 & 0.7381 & 0.8571 & \textbf{0.9167} & 0.008** & 0.003** \\
Recall@3 & 0.1012 & 0.1369 & \textbf{0.1440} & 0.015* & 0.009** \\
Precision@3 & 0.3373 & 0.4563 & \textbf{0.4802} & 0.007** & 0.004** \\
NDCG@3 & 0.3679 & 0.4832 & \textbf{0.5287} & 0.005** & 0.001*** \\
\midrule
Hit Ratio@5 & 0.9048 & 0.9286 & \textbf{0.9524} & 0.042* & 0.015* \\
Recall@5 & 0.1714 & \textbf{0.1964} & 0.1893 & 0.028* & 0.156 \\
Precision@5 & 0.3429 & \textbf{0.3929} & 0.3786 & 0.033* & 0.245 \\
NDCG@5 & 0.3637 & 0.4314 & \textbf{0.6394} & 0.018* & 0.002** \\
\midrule
Hit Ratio@10 & \textbf{1.0000} & 0.9762 & \textbf{1.0000} & 0.045* & - \\
Recall@10 & 0.3298 & 0.2964 & \textbf{0.3310} & 0.039* & 0.267 \\
Precision@10 & 0.3298 & 0.2964 & \textbf{0.3310} & 0.042* & 0.334 \\
NDCG@10 & 0.3468 & 0.3503 & \textbf{0.4528} & 0.156 & 0.008** \\
\bottomrule
\end{tabular} \\[2ex]
\end{subtable}

\begin{subtable}{\textwidth}
\centering
\caption{Item-based CF}
\label{tab:item_based}
\begin{tabular}{lccccc}
\toprule
\multirow{2}{*}{\textbf{Metrics}} & \multirow{2}{*}{\textbf{Non-Ranker}} & \multirow{2}{*}{\textbf{Zero-shot}} & \multirow{2}{*}{\textbf{SFT-DPO}} & \multicolumn{2}{c}{\textbf{p-value}} \\
\cmidrule[\lightrulewidth](lr){5-6}
& & & & \textbf{Zero vs Non} & \textbf{SFT vs Non} \\
\midrule
Hit Ratio@3 & \textbf{0.499} & 0.468 & 0.484 & 0.823 & 0.752 \\
Recall@3 & 0.0468 & 0.0475 & \textbf{0.049} & 0.823 & 0.538 \\
Precision@3 & \textbf{0.2127} & 0.1967 & 0.202 & 0.653 & 0.544 \\
NDCG@3 & 0.2081 & 0.2035 & \textbf{0.2106} & 0.954 & 0.755 \\
\midrule
Hit Ratio@5 & 0.622 & 0.640 & \textbf{0.649} & 0.652 & 0.223 \\
Recall@5 & 0.0755 & 0.0671 & \textbf{0.0816} & 0.468 & 0.047* \\
Precision@5 & 0.1882 & 0.170 & \textbf{0.2016} & 0.531 & 0.027* \\
NDCG@5 & 0.1957 & 0.1856 & \textbf{0.2052} & 0.400 & 0.142 \\
\midrule
Hit Ratio@10 & 0.820 & 0.815 & \textbf{0.872} & 0.729 & 0.016* \\
Recall@10 & 0.1338 & 0.137 & \textbf{0.1652} & 0.872 & 0.000* \\
Precision@10 & 0.166 & 0.1696 & \textbf{0.2036} & 0.876 & 0.000* \\
NDCG@10 & 0.1773 & 0.1804 & \textbf{0.2056} & 0.664 & 0.001* \\
\bottomrule
\end{tabular}
\end{subtable}

\begin{subtable}{\textwidth}
\centering
\caption{Knowledge Graph}
\label{tab:knowledge_graph}
\begin{tabular}{lccccc}
\toprule
\multirow{2}{*}{\textbf{Metrics}} & \multirow{2}{*}{\textbf{Non-Ranker}} & \multirow{2}{*}{\textbf{Zero-shot}} & \multirow{2}{*}{\textbf{SFT-DPO}} & \multicolumn{2}{c}{\textbf{p-value}} \\
\cmidrule[\lightrulewidth](lr){5-6}
& & & & \textbf{Zero vs Non} & \textbf{SFT vs Non} \\
\midrule
Hit Ratio@3 & \textbf{0.499} & 0.468 & 0.484 & 0.2626 & 0.1823 \\
Recall@3 & \textbf{0.0517} & 0.0475 & 0.049 & 0.5132 & 0.245 \\
Precision@3 & \textbf{0.2127} & 0.1967 & 0.202 & 0.5033 & 0.163 \\
NDCG@3 & \textbf{0.2081} & 0.2035 & 0.2066 & 0.3779 & 0.1367 \\
\midrule
Hit Ratio@5 & \textbf{0.663} & 0.622 & 0.649 & 0.7544 & 0.156 \\
Recall@5 & 0.0816 & 0.0755 & \textbf{0.0841} & 0.9542 & 0.189 \\
Precision@5 & \textbf{0.2078} & 0.1882 & 0.2016 & 0.9365 & 0.245 \\
NDCG@5 & 0.2052 & 0.1957 & \textbf{0.2062} & 0.6213 & 0.334 \\
\midrule
Hit Ratio@10 & \textbf{0.876} & 0.815 & 0.872 & 0.4703 & 0.156 \\
Recall@10 & \textbf{0.1665} & 0.137 & 0.1652 & 0.7348 & 0.267 \\
Precision@10 & \textbf{0.2051} & 0.1696 & 0.2036 & 0.6764 & 0.189 \\
NDCG@10 & 2.046 & 0.1804 & \textbf{0.2056} & 0.5626 & 0.245 \\
\bottomrule
\end{tabular} \\[2ex]
\end{subtable}
\end{table*}

In terms of explainability, the SFT-DPO model demonstrates significant improvements over the zero-shot baseline. A human evaluation with 5 participants, each reviewing 20 samples from both models with shuffled presentation order, rated the clarity and persuasiveness of explanations on a 5-point Likert scale. The SFT-DPO model achieved an average score of \textbf{4.3}, compared to \textbf{3.6} for the zero-shot baseline, indicating better explainability. A paired t-test confirmed the statistical significance of this improvement (\textit{p-value} < \textbf{0.01}). While the zero-shot model primarily relied on movie genres for explanations, the SFT-DPO model provided more comprehensive explanations by integrating movie overviews and user history summaries, as detailed in Appendix ~\ref{tab:sample_comparison}.

\section{Analysis}
The effectiveness of the LLM Re-Ranker largely depends on the quality of the original recommendation model. For models with poor ranking performance (e.g., randomly ranked items or relatively weak models like item-based collaborative filtering), the LLM Re-Ranker demonstrates significant improvements. However, for stronger recommendation models, while the LLM Re-Ranker can still enhance performance to some extent, the improvement is relatively limited. For instance, the observed p-value remains around 0.1–0.2, failing to reach the commonly accepted significance threshold of 0.05. Furthermore, we observed a performance degeneration phenomenon when the base recommendation model is exceptionally strong, as seen in the case of knowledge graph-based systems. This may be because knowledge graph-based models incorporate richer contextual information, such as graph links between users and items, as well as detailed text descriptions or item features—information that may not be fully utilized by the LLM Re-Ranker.

After training, we noted an improvement in the reasoning and generation abilities of the LLM. The trained LLM could effectively summarize a user's profile based on their interaction history and provide high-level, compelling reasons for recommendations. In contrast, the zero-shot LLM often focused narrowly on a single movie the user had previously watched, failing to deliver a clear or persuasive rationale for its recommendations.

\section{Limitations and Future Improvement}
As discussed in the previous section, the LLM Re-Ranker struggles to fully leverage graph structure information, which likely explains its limited ability to significantly enhance the accuracy of advanced recommendation models. To address this, a potential improvement involves integrating graph information into the LLM. One possible solution is to extract a subgraph from the prediction model, train a graph encoder, and incorporate a mapping layer to transform the graph information into vector representations. These vectors could then be used as prefix tokens for the LLM, as explored in \cite{he2024g}.

Additionally, constructing a higher-quality dataset remains an important area for improvement. Evaluating the quality of a generated dataset is inherently challenging. In this study, we conducted a simple experiment to test for significant differences between the generated datasets and a small set of manually created entries. However, the limited number of manually curated samples may not provide compelling evidence to fully validate the dataset's quality. Future work could focus on developing a more sophisticated methodology for constructing and evaluating training datasets to ensure their alignment with human expectations and task requirements.

\section{Conclusion}
This work introduces an approach to integrating LLM into recommendation systems, addressing challenges such as explainability and popularity bias. By combining traditional recommendation models with an LLM-based re-ranker, we improved ranking accuracy and interpretability. After two-stage training, our model outperformed zero-shot baselines and demonstrated significant improvements in metrics like NDCG. Additionally, to mitigate internal biases, we employed bootstrapping and a self-consistency module, ensuring robust and fair rankings. Human evaluations confirmed the alignment of LLM-generated datasets with human expectations and demonstrated the trained model’s ability to provide more compelling explanations. While limitations remain in leveraging graph-based information and dataset construction, this study highlights the potential of LLMs as explainable re-rankers, paving the way for more transparent and user-centric recommendation systems.

\bibliography{main}

\appendix

\section{Appendix A: Sample Comparison Between Zero-Shot and SFT-DPO Models}
\label{sec:appendix}

Table \ref{tab:sample_comparison} provides six samples of the explanations generated by the zero-shot and SFT-DPO models, along with their respective ratings from human evaluators. These samples illustrate the differences in the comprehensiveness and persuasiveness of the explanations provided by the two models.

\begin{table*}[h]
\centering
\caption{Comparison of Explanations and Ratings Between Zero-Shot and SFT-DPO Models}
\label{tab:sample_comparison}
\begin{tabular}{|p{5cm}|c|p{5cm}|c|}
\hline
\textbf{Zero-Shot Model Explanation}                                                                                                   & \textbf{Rating} & \textbf{SFT-DPO Model Explanation}                                                                                                   & \textbf{Rating} \\ \hline
"Forrest Gump - Reason: User's historical interactions include multiple romantic and drama movies, Forrest Gump is a romantic drama."                                & 3             & "The blend of adventure, romance, and humor resonates with the user's enjoyment of light-hearted yet meaningful narratives."      & 4             \\ \hline
"User's historical interactions include Adventure and Comedy genres. Monty Python and the Holy Grail is a well-known comedy adventure film."                                       & 3             & "The psychological depth and mystery elements align with the user's interest in thought-provoking narratives." & 4             \\ \hline
"Similar to "Indiana Jones and the Last Crusade", "Raiders of the Lost Ark" is an adventure and action movie that the user has not yet seen."                                       & 4             & "This movie explores the complexities of relationships and the challenges of love, aligning with the user's interest in romantic dramas." & 5            \\ \hline

\end{tabular}
\end{table*}

The zero-shot model explanations typically rely on general genre matching, which lacks depth and specificity. In contrast, the SFT-DPO model combines genre information with contextual elements such as movie overviews and user history, resulting in explanations that are more comprehensive and persuasive. This difference in approach is reflected in the higher ratings given to the SFT-DPO model's explanations.

\end{document}